\documentclass[aps,prd,amsmath,amssymb,twocolumn,superscriptaddress,preprintnumbers,nofootinbib]{revtex4-1}
\pdfoutput=1
\usepackage{graphicx}
\usepackage{amsthm}
\usepackage{amsmath}
\usepackage{graphicx,subfigure}%
\usepackage{dcolumn}%
\usepackage{bm}%
\usepackage{slashed}

\newcommand{\be}{\begin{eqnarray}}
\newcommand{\ee}{\end{eqnarray}}

\DeclareMathOperator{\erf}{Erf}

\newcommand{\keV}{\ensuremath{\mathrm{keV}}}
\newcommand{\MeV}{\ensuremath{\mathrm{MeV}}}
\newcommand{\GeV}{\ensuremath{\mathrm{GeV}}}

\newcommand{\kg}{\ensuremath{\mathrm{kg}}}

\newcommand{\keVee}{\ensuremath {\mathrm{keV_{\!ee}}}}

\newcommand{\Er}{\ensuremath {E_R}}

\newcommand{\cpd}{\ensuremath {\mathrm{cpd}}}

\renewcommand{\day}{\ensuremath {\mathrm{day}}}

\newcommand{\Kforty}{\ensuremath{{}^{40}\mathrm{K}}}

\newcommand{\Arforty}{\ensuremath{{}^{40}\mathrm{Ar}}}
\newcommand{\Caforty}{\ensuremath{{}^{40}\mathrm{Ca}}}
\newcommand{\Thalf}{\ensuremath{T_{1/2}}}

\newcommand{\BREC}{\rm BR_{EC}}

\begin{document}

\title{On an unverified nuclear decay and its role in the DAMA experiment}
\author{Josef Pradler}
\email{jpradler@pha.jhu.edu}
\affiliation{Department of Physics and Astronomy, Johns Hopkins University, Baltimore, MD 21218, USA}
\author{Balraj Singh}
\email{ndgroup@mcmaster.ca}
\affiliation{Department of Physics \& Astronomy, McMaster University
1280 Main St. W, Hamilton, Ontario, Canada, L8S 4L8}
\author{Itay Yavin}
\email{iyavin@perimeterinstitute.ca}
\affiliation{Perimeter Institute for Theoretical Physics 31 Caroline St. N, Waterloo, Ontario, Canada N2L 2Y5.}
\affiliation{Department of Physics \& Astronomy, McMaster University
1280 Main St. W, Hamilton, Ontario, Canada, L8S 4L8}

\begin{abstract}
  The rate of the direct decay of \Kforty\ to the ground state of
  \Arforty\ through electron capture has not been experimentally
  reported. Aside from its inherent importance for the theory of
  electron capture as the only such decay known of its type (unique
  third-forbidden), this decay presents an irreducible background in
  the DAMA experiment. 
 We find that the presence of this background, as well as others,
 poses a challenge to any interpretation of the
 DAMA results in terms of a Dark Matter model with a small
 modulation
 fraction. A 10~ppb contamination of
 natural potassium requires a 20\% modulation fraction or more. A
 20~ppb contamination, which is reported as an upper limit by DAMA,
 disfavors any Dark Matter origin of the signal. This conclusion is based on the efficiency of detecting \Kforty\ decays as inferred from simulation. We propose measures to help clarify the
 situation.
\end{abstract}

\pacs{12.60.Jv, 12.60.Cn, 12.60.Fr}
\maketitle

\section{Introduction}

A $\Kforty$ concentration of $0.0117(1)\%$~\cite{DDEPWG} in naturally occurring
potassium, $^{\mathrm{nat}}$K, is an ubiquitous source of
radioactivity.
Its decays in solidified rocks are responsible for the air's argon
content and in a human body it is the element with one of the largest
activities.
The chemical similarity of sodium and potassium is not only
responsible for a per mille fraction of $^{\mathrm{nat}}$K in seawater
but may ultimately be the reason why trace amounts of potassium can
also be found in ultra-radiopure scintillating crystals grown from NaI
powders. These crystals are employed in rare event searches, such as
in Dark Matter (DM) direct detection experiments where radioactive
backgrounds pose one of biggest challenges limiting sensitivity.

\begin{figure}[tb]
\begin{center}
\includegraphics[width=1\columnwidth]{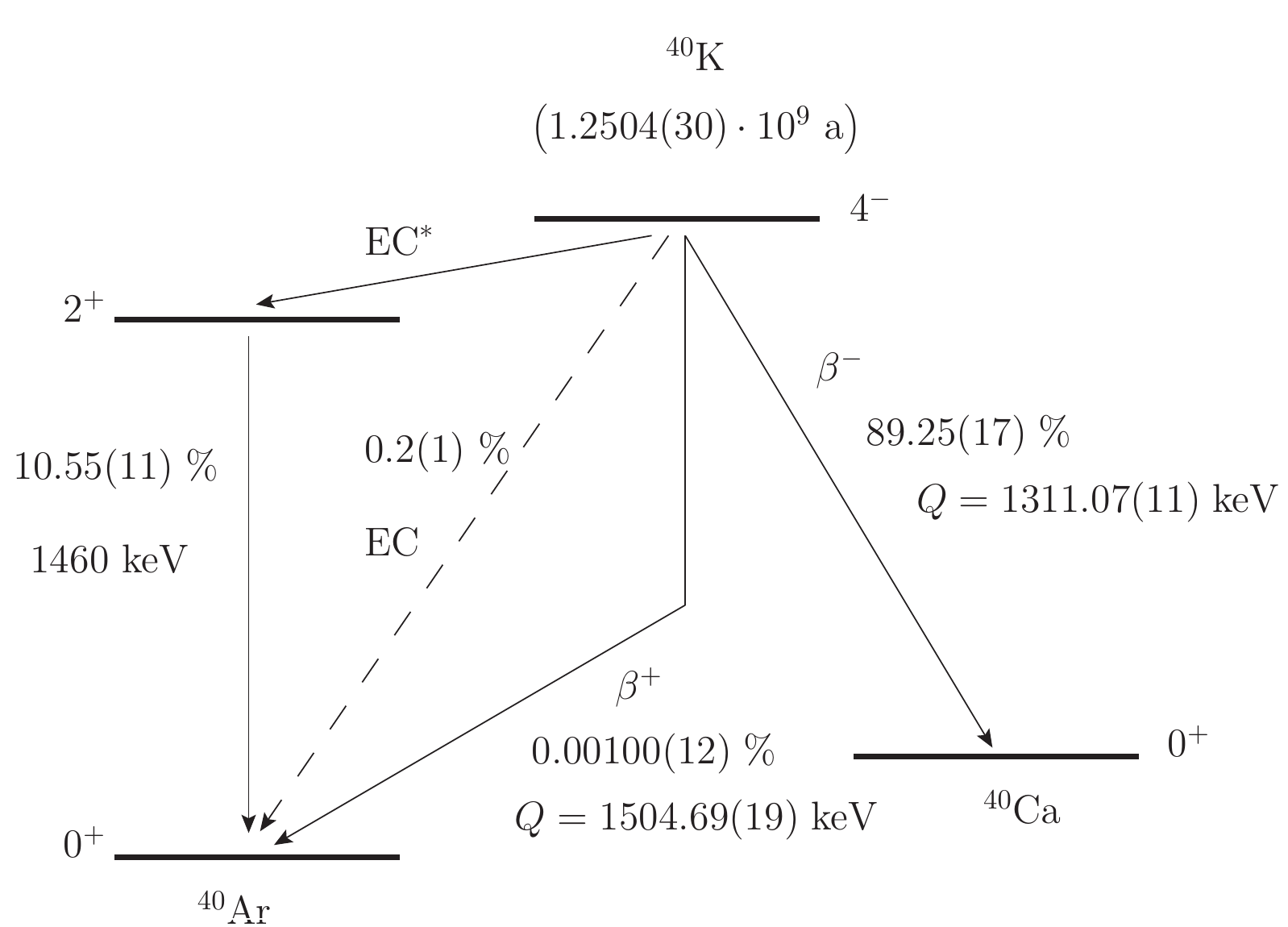}
\end{center}
\caption{Decay scheme of $\Kforty$ with numbers based on~\cite{DDEPWG}. 
 The dashed line depicts the direct decay to the ground state of
 $\Arforty$ for which no dedicated measurement  has been reported. The
 branching ratio quoted is extracted from the measured $\beta^+$ decay
 rate and the extrapolated theoretical ratio ${\rm EC}/\beta^+$.}
\label{fig:decay_scheme}
\end{figure}

The decay scheme of $\Kforty$ is shown in Fig.~\ref{fig:decay_scheme}.
All kinematically available energy levels in \Arforty\ and \Caforty\
are populated in the \Kforty\ decay:
\begin{enumerate}
\item The dominant mode is the $\beta^-$ decay to the calcium ground
  state with a half-life $\Thalf(\beta^{-}) = 1.407(7)\times 10^9$~yr
  and an endpoint energy of $Q_{\beta^{-}} = 1311.07(11)~\keV$.
\item The argon branch is dominated by electron capture (EC) into the
  excited state $\Arforty^{*}(1460\,\keV)$ with $T_{1/2}({\rm
    EC^*,1460}) = 11.90(11)\times 10^9$~yr. The nucleus subsequently
  de-excites by emission of a $\gamma$-ray.
\item Positron emission to the ground state of $\Arforty$ is
  possible. It has been measured~\cite{Engelkemeir:1962} and found to
  have a small branching ${\rm BR}_{\beta^{+}} = 0.00100(12)\%$.
\item Last but not least is EC directly into the ground state of
  \Arforty.  Little is known empirically about this decay and the
  current branching quoted in the literature, ${\rm BR}_{\mathrm{EC}}
  = 0.2(1)\%$~\cite{DDEPWG}, is based on the measurement of ${\rm
    BR}_{\beta^{+}} $ and a theoretical extrapolation from the EC to
  positron ratio for 1$^{\rm st}$ and 2$^{\rm nd}$ forbidden-unique
  transitions.

\end{enumerate}

Experimentally, the most accessible modes are (1) and (2) because the
respective electron or gamma-ray are readily detectable and numerous
measurements exist. The quoted numbers are world-averages obtained
in~\cite{be2004table}.
The observation of positron emission (3) is challenging because of its
low intensity and pair-production background, but it was first
reported in~\cite{Engelkemeir:1962}.  In contrast to the above modes,
the rare EC to the ground state of $\Arforty$ in (4) has never been
reported from a dedicated measurement. Since it is a transition from
$J^{\pi}(\Kforty)=4^-$ to $J^{\pi}(\Arforty)=0^+$ it is classified as
\textit{unique third-forbidden} (3U) and it is the only such EC decay
known~\cite{NSR1998SI17, nndc-ensdf}. As such, its empirical
verification would bring an important closure to the theory of unique
forbidden decays.

The energy release involved in this special decay is carried almost
entirely by the emitted neutrino. Only about 3 keV of energy is
observable from the Auger electron emission and the X-ray yield upon
K-shell capture. As such, this decay forms an important irreducible
background to the DAMA~\cite{Bernabei:2010mq} experiment which looks
for DM scattering in the laboratory. Since the experiment does not
employ any discrimination between electron/gamma and nuclear-recoil
activity, the release of only 3 keV associated with the rare EC decay
of \Kforty\ shows up directly in the region of interest for dark
matter.

The purpose of this letter is to point out that---despite the
omnipresent nature of \Kforty---a dedicated measurement of one of its
decay modes, namely the EC directly to the ground state of \Arforty ,
remains outstanding. We begin by discussing what is currently known
about this decay, carefully separating what is empirically established
from what is based on a theoretical extraction. We then discuss the
importance of this background in the DAMA experiment and how it
affects the interpretation of the claimed results. We close with a
brief discussion of how this special decay may be measured directly.

\section{\boldmath$\Kforty$ rare EC decay}
\label{sec:experimental-status-1}

Given the absence of a reported measurement for the EC mode, its
branching ratio, ${\rm BR}_{\mathrm{EC}}$, can be inferred from the
theoretical ratio ${\rm BR}_{\mathrm{EC}}/{\rm BR}_{\beta^{+}}$.  The
branching ratio of $\beta^{+}$ decay, ${\rm BR}_{\beta^{+}}$, has been
measured directly.
The rational behind this is that to leading order the atomic and
nuclear pieces factorize and the branchings ${\rm
  BR}_{\mathrm{EC}}$ and ${\rm BR}_{\beta^{+}}$ are dominated by the
same nuclear matrix element, which then cancels in the ratio. Hence,
the calculation of the ratio ${\rm BR}_{\mathrm{EC}}/{\rm
  BR}_{\beta^{+}}$ becomes essentially a question of accurately
modeling the atomic wave functions of the electron/positron and the
neutrino involved.

Explicitly, the rate for positron emission is given by
\begin{align}
 \label{eq:lambdabplus}
 \lambda_{\beta^{+}} \equiv \frac{\ln{2}}{T_{1/2}(\beta^{+})} =
 \frac{g^2}{2\pi^3} \int_0^{p_0} dp\, p^2 p_{\nu}^2
 C_{3U}(Z_{\mathrm{Ar}},E) ,
\end{align}
where $g$ is the Fermi constant times the cosine of the Cabibbo angle;  $p\,(E)$ is the positron momentum (energy) with endpoint $p_0\,(E_0)$; the
neutrino momentum is $p_{\nu} = E_0-E$.
The shape factor $C_{3U}$ contains the nuclear matrix element
$\mathcal{M}_{3U}$ and the Coulomb correction in the form of the Fermi
function $F(Z,E)$.
The neutrino and positron are predominantly emitted in the triplet state
allowing for one unit less in lepton orbital angular momentum when
mediating a nuclear spin change of $J=4$. Denoting by $j_{\nu}$ and
$j_e$ their respective angular momenta,
\begin{align}
  C_{3U} = g_A^2|\mathcal{M}_{3U}|^2 R^6 \times \sum_{ j_{\nu},j_e } \text{(atomic part)}_{j_{\nu} , j_e},
\end{align}
where the sum is subject to the condition $ j_{\nu} + j_e = J$; $g_A$
is the axial vector coupling and $R$ denotes the nuclear radius. We
follow \cite{Gove:1971} in the evaluation of the atomic part.

Turning to EC, this process mainly occurs through K-shell capture and the rate
is to a good approximation given by%
\footnote{Higher shell captures contribute only about 10\% to the
  total rate and yield energy depositions below the DAMA
  threshold. The error in the ratio induced by neglecting those
  contributions is much smaller than the overall uncertainty in this quantity.
}
\begin{align}
\label{eq:lambdaEC}
\lambda_{\mathrm{EC}} \equiv \frac{\ln{2}}{T_{1/2}(\mathrm{EC})} = \frac{g^2}{2\pi^3} f_K  C_K ,
\end{align}
where $f_K$ contains the 1S amplitude for the bound state radial electron
wave function and 
\begin{align}
\label{eq:CK}
  C_K = \frac{q_K^6 R^6}{11025} \times g_A^2|\mathcal{M}_{3U}|^2 .
\end{align}
Here $q_K$ is the momentum of the emitted neutrino. Equations
(\ref{eq:lambdabplus}) to (\ref{eq:CK}) show that the nuclear
matrix element and the strong sensitivity to the nuclear radius cancel
in the ratio $\lambda_{\mathrm{EC}}  /\lambda_{\beta^{+}} =
{\rm BR}_{\mathrm{EC}} /{\rm BR}_{\beta^{+}} $.
Using Eqs.~(\ref{eq:lambdabplus}) to (\ref{eq:CK}) directly we 
find a value ${\rm BR}_{\mathrm{EC}}/{\rm BR}_{\beta^{+}} = 190$ when
using the atomic data collected in~\cite{Bambynek:1977zz}. When using the approximations to the shape factor given
in~\cite{Davidson:1951zz} we find a slightly smaller value of ${\rm BR}_{\mathrm{EC}}/{\rm BR}_{\beta^{+}} =150$. 

Early evaluations~\cite{Engelkemeir:1962} used $\lambda_{\rm EC}/\lambda_{\beta^{+}} = 155$ without further reference to literature. The most recent one~\cite{DDEPWG} uses
\begin{align}
\label{eq:ECtoBetaPlus}
  \frac{\lambda_{\mathrm{EC}}}{\lambda_{\beta^{+}}} =200(100)
\quad \Rightarrow \quad  {\rm BR}_{\mathrm{EC}} = 0.2(1)~\%,
\end{align}
This is in good agreement
with what we find using the direct calculation above, however it is important to note that this latter ratio is an extrapolated
theoretical expectation and not computed from the above equations.
The \texttt{LOGFT} computer program~\cite{nndc-logft} used in nuclear
data evaluations cannot compute the EC/$\beta^+$ for the 3U ratio in
(\ref{eq:ECtoBetaPlus}). Instead the ratios to unique first- (1U) and
second-forbidden (2U) transitions are calculated, ${\rm
  BR}_{\mathrm{EC}}/{\rm BR}_{1U} = 8.51(9) $ and ${\rm
  BR}_{\mathrm{EC}}/{\rm BR}_{1U} = 45.20(47) $,
respectively. Assuming a constant increase by the same factor, ${\rm
  BR}_{\mathrm{EC}}/{\rm BR}_{3U} = {\rm BR}_{\mathrm{EC}}/{\rm
  BR}_{\beta^{+}} = 240 $ is obtained from which the value
in~(\ref{eq:ECtoBetaPlus}) has been adopted.

There is no doubt of the success of leading order theoretical
predictions based on the \mbox{V-A} theory of weak interactions in
explaining weak decay strengths and ratios observed throughout the
periodic table. Yet, the $\Kforty$ decays to the ground states of
\Arforty\ and \Caforty\ are the only known 3U transitions realized in
nature~\cite{NSR1998SI17, nndc-ensdf}. Hence, for the EC in question
it is not obvious how to truly ensure the validity of
(\ref{eq:ECtoBetaPlus}), \textit{e.g.} by gauging it against
measurements of other 3U transitions. Theoretical attempts to match
the calculated $\beta^{+}$ and $\beta^{-}$ strengths to actual
measurements yield discrepancies by a factor $\sim 3$ for
$\Kforty(\beta^-)$ to $^{40}{\rm Ca}$ and a factor $\sim 6$ for
$\Kforty(\beta^+)$ to $\Arforty$ when simple nuclear shell models are
employed~\cite{Warburton:1970}. These discrepancies are very likely the result of
inaccuracies in the evaluation of the corresponding nuclear matrix
elements (which cancel when forming the ratio), but they make it
difficult to feel confident about the theoretical estimation in
Eq.~(\ref{eq:ECtoBetaPlus}). At the very least, they remind us of the
importance of an empirical verification.

It is also possible to obtain a \textit{theory-independent} estimate of ${\rm BR}_{\mathrm{EC}}$
from measurements of the total half-life of $\Kforty$.  Indeed, this
is facilitated by a recent high-precision
measurement~\cite{Kossert:2004}, $T_{1/2} = 1.248(9)\times 10^9$~yr.%
\footnote{
  Uncertainties in the decay scheme affect the error (but not the
  measured value) of $\Thalf$. This is because $\beta$-decay and EC are not
  registered with equal efficiencies. The error on
  ${\rm BR}_{\mathrm{EC}}$ in (\ref{eq:ECgs-possible}) is calculated
  self-consistently which is also reflected by a larger adopted error
  on $\Thalf$ than what is originally quoted in~\cite{Kossert:2004}.}
Neglecting positron emission we can write,
\begin{align*}
  T^{-1}_{1/2}\left({\rm EC} \right) =T^{-1}_{1/2} -
  \left[T^{-1}_{1/2}(\beta^-)+T^{-1}_{1/2}({\rm EC^*,1460}) \right] ,
\end{align*}
and using the measured values for the $\beta^-$ and EC$^*$ branches one obtains,
\begin{align}
\label{eq:ECgs-possible}
  {\rm BR}_{\mathrm{EC}} \equiv \frac{ T_{1/2} }{T_{1/2}\left({\rm EC} \right)} = 0.8(8)~\% \, .
\end{align}
Given the large uncertainties it is clear that the available lifetime measurements are currently not sensitive enough to
pin down the EC strength with reasonable accuracy. This special branch thus deserves its own dedicated measurement. Aside from its inherent significance, the precise activity of this branch also forms an important background in the DAMA experiment as we discuss in the following section.

\section{Potassium background in DAMA}
\label{sec:bkg}

The DAMA experiment is situated in the Gran Sasso underground
laboratory and searches for DM scattering off nuclei with a target made of 25 NaI(Tl)
crystals amounting to almost 250~\kg\ in
mass~\cite{Bernabei:2008yh}. The detector registers energy depositions
between $2~\keV$ to tens of MeV of almost any source: electrons,
gamma- and X-rays, muons, alpha particles, and nuclear recoils. Events
of the last class with recoil energies below a few $\keVee$%
\footnote{Electron equivalent energy in keV.}
are expected from the scattering of DM particles in the galactic halo
against the target nuclei if the DM mass is $\gtrsim
10~\GeV$. Importantly, no discrimination procedure to veto against the
other types of recoils is employed.  Instead, the most compelling
feature of the reported result is the annually modulating
``single-hit''\footnote{``Single-hit'' means that a prospective event
  must pass a coincidence veto with all other
  crystals. ``Multiple-hit'' events can, \textit{e.g.}, be induced by
  $\gamma$-radiation accompanied by radioactive decay.} event rate
between $2-6\,\keVee$ with a phase that is compatible with the phase
expected from DM (June~2).  The collaboration reports the
\textit{residual recoil rate} from which the average count rate of a
cycle was subtracted.%
\footnote{A DAMA cycle is very roughly one year but is subject to
  variations which are not further detailed by the collaboration.}
The residual rate exhibits an annually modulating pattern that can be
nicely fitted to the model $S_m \cos{\omega (t-t_0)}$ with $\omega =
2\pi/(1\,\mathrm{yr})$, a modulation amplitude of $S_m\simeq
0.04\,\cpd/\kg$ in the $2-4\,\keV$ bin and a phase of $t_0 \approx
140$ days. This is compatible with the expectation from a Maxwellian
DM halo velocity distribution (see for example the recent
review~\cite{Freese:2012xd}).

However, knowledge of the modulated rate, $S_m$, without a detailed
insight into the unmodulated rate, $R_0$, makes the interpretation of
the signal in terms of DM difficult. What is reported is that $R_0
\sim 1\,\cpd/\kg/\keV$ for the spectrum below 10\,\keV. Essentially
no information is available as to what it is comprised of. Radioactive
backgrounds in rare event searches arise predominantly from naturally
occurring radioactive isotopes (such as \Kforty) in the detector
material and its surroundings, from cosmogenically activated elements
(\textit{e.g.}  $^{129}$I), and from elements in the natural uranium
($^{238}$U) and thorium ($^{232}$Th) decay chains.

DAMA addresses all those background sources in~\cite{Bernabei:2008yh}
where they quote the concentrations of the different radio-isotopes. The main
shortcoming of this study is that the influence on the low energy
single-hit spectrum is not addressed quantitatively.
The observed rate below 10\,\keV\ is essentially flat, with the
exception of a bump around $3\,\keV$.
Flat spectra are typical for beta decays due to the Coulomb
corrections at small electron emission energies as discussed below.  Bumps in the keV
energy region point towards EC. It has been speculated for a long time
that the bump seen at  3 keV may very well be associated with
K-shell electron capture of $\Kforty$. The nuclear recoil in this
decay is negligible and the scintillator detects the entire K-shell
electron binding energy of \Arforty\ ($3.2\,\keV$) released in the form
of X-rays and Auger electrons.
It is important to note that the position of the peak coincides with
the maximum of $S_{m}$. Although the DAMA collaboration insists that
these two bumps are of different origin, the presence of a poorly
understood background in the signal region is at least unsettling.

\begin{figure}[tb]
\begin{center}
\includegraphics[width=1\columnwidth]{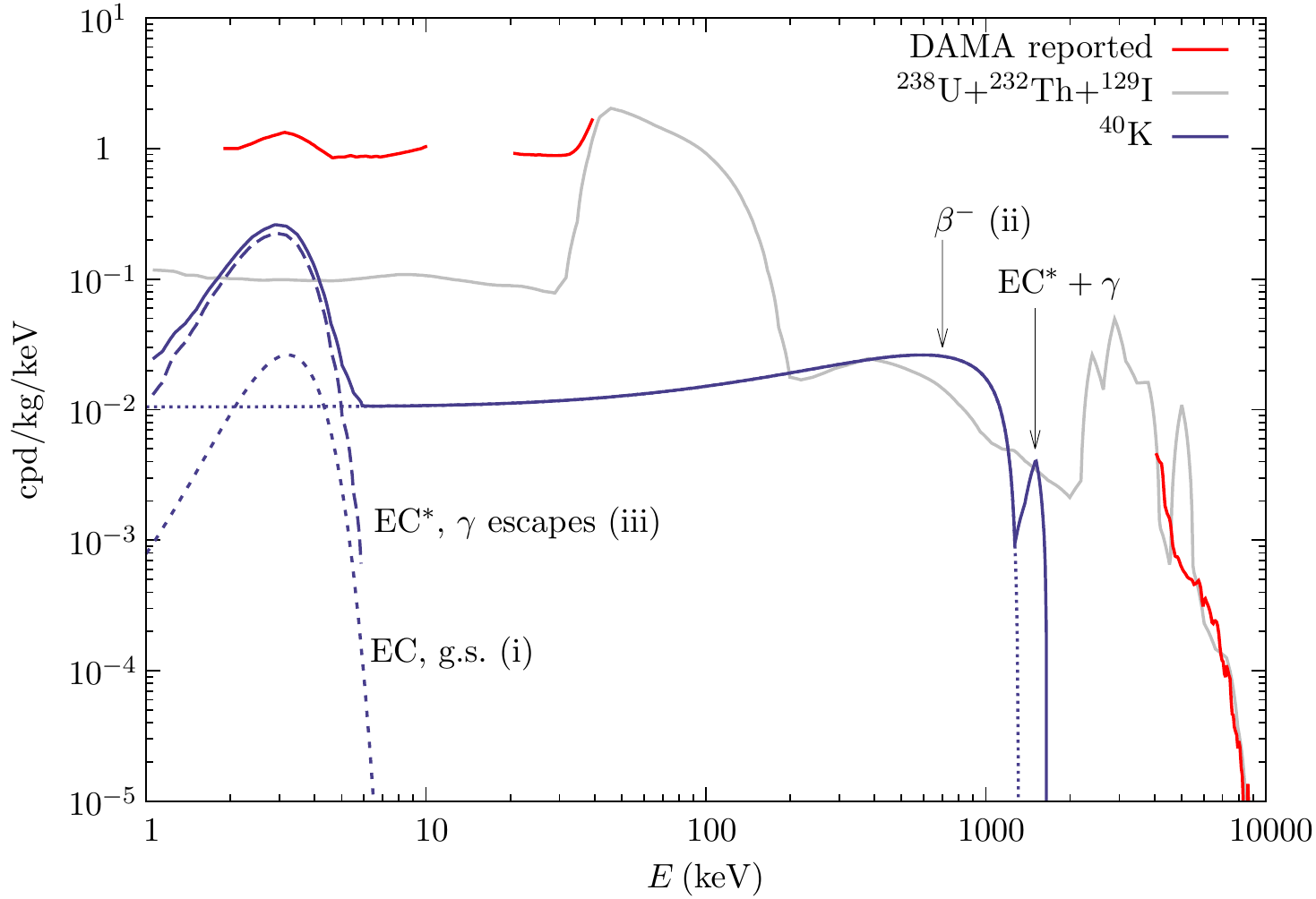}
\end{center}
\caption{DAMA ``single-hit'' spectrum and simulated backgrounds. The
  three disjont pieces (in red) are DAMA reported rates from
  \cite{Bernabei:2008yh} and \cite{Bernabei:2009zzb}. The statistical
  uncertainty in the signal region is negligible. The gray line is the
  spectrum from $^{129}$I, $^{238}$U, and $^{232}$Th obtained in
  \cite{Kudryavtsev:2010zza} with respective concentrations 0.2~ppt,
  5~ppt, and 1.7~ppt. The blue lines show the various $\Kforty$
  contributions as detailed in the text with a $^{\mathrm{nat}}$K
  contamination level of $c_K = 10$~ppb. The highest energy
  $\alpha$-peak in the gray line is vetoed by the DAMA DAQ
  system~\cite{Bernabei:2008yh}.}
\label{fig:dama_low_energy_spectrum}
\end{figure}

Based on the reported levels of radioactivity, the authors
of~\cite{Kudryavtsev:2010zza} were the first to try and evaluate the
resulting rate in the DM signal region.  Their main conclusion was
that the unmodulated rate seems to require higher than reported
concentrations of some of the isotopes. 
Moreover, in-situ contaminations of the crystals were found to
dominate the low-energy spectrum as external radio-impurities are
limited by the prominent back-scatter peak they produce for $\gtrsim
40\,\keV$.
Fig.~\ref{fig:dama_low_energy_spectrum} shows the unmodulated
``single-hit'' spectrum: the three (red) line segments are reported
event rates by DAMA~\cite{Bernabei:2008yh}. The gray line is taken
from~\cite{Kudryavtsev:2010zza} and shows the simulated spectrum from
$^{238}$U, $^{232}$Th, and $^{129}$I in-situ decays. The broad peak between
$40-100\,\keV$ is from $^{129}$I with little contribution elsewhere.
Hence, the low-energy count rate is dominated by $^{238}$U and
$^{232}$Th resulting in an essentially flat spectrum below 10\,\keV.
The (quenched) $\alpha$-decays in those chains result in peaks above
$2\,\MeV$, which are used by DAMA to derive the reported
concentrations, but those are found to be insufficient to explain the
observed rate below $10\,\keV$. Though the contributed amount from
$^{238}$U and $^{232}$Th at low energies is limited by the observation
of the spectrum at high energies ($\gtrsim$ few MeV), it is important
to recognize that the contribution from \Kforty\ decay is not
similarly constrained.

DAMA reports an upper limit on the contamination level of
$^{\mathrm{nat}}$K as $c_{\mathrm{K}} \leq
20$~ppb~\cite{Bernabei:2008yh}.
They do so by using the decay of \Kforty\ to the excited state of
\Arforty\ to look for double-coincidence events where the 3 keV is
registered in one crystal and the 1460 keV gamma-ray is registered
entirely in another crystal. The contamination level is obtained by
dividing the observed double-coincidence rate by the probability for
such events to occur. The latter was inferred from Monte Carlo (MC)
simulation but no detailed description of this procedure has been
provided. Given the indirect nature of the $^{\mathrm{nat}}$K
determination and the lack of knowledge of potential systematic
uncertainties, it may well be possible that the potassium
contamination is larger than what is reported.

\Kforty\ contributes to the low energy spectrum in three distinct
ways:
\begin{enumerate}
\item[i)] The direct EC decay to the ground state of \Arforty, which is
  the principal subject of this paper, contributes solely to the bump
  at 3.2~keV.
\item[ii)] The electron emission associated with the \Kforty\ to
  \Caforty\ decay contributes a flat background that extends all the
  way up to the end-point energy of 1311 keV.
\item[iii)] The EC decay to the excited state of \Arforty\ results in
  the same low energy contribution to the 3~keV bump, but it is
  followed by the emission of 1460~keV gamma-ray. This decay
  contributes to the single-hit rate at low energy only when the
  1460~keV gamma-ray escapes undetected.
\end{enumerate}

The direct decay (i) is the easiest to calculate since it only depends
on the contamination level of $^{\mathrm{nat}}$K in the NaI crystals
and the branching ratio of this decay. If we assume a contamination
level of $c_{\mathrm{K}}$ then the activity from $\Kforty$ decays to
the ground state is given by,
\begin{align}
  \Gamma_{\mathrm{EC}} & = c_{40} c_{\mathrm{K}} {\rm BR}_{\mathrm{EC}}
   \frac{ N_A \ln 2}{A_{\mathrm{K}} \Thalf M_u}, \\ \nonumber
  &= \frac{0.11}{\kg_{\mathrm{NaI}}\,\day} \left( \frac{{\rm BR}_{\mathrm{EC}}}{0.2\%}
  \right) \left( \frac{c_{\mathrm{K}}}{20\,\mathrm{ppb}} \right)
\end{align}
Here $N_A$ is Avogadro's number, $M_u=1\,\mathrm{g}/\mathrm{mol}$ is
the molar mass constant, $\Thalf$ is the total lifetime of $\Kforty$,
$c_{40} = 0.0117(1)\%$ is the $\Kforty$ fraction in
$^{\mathrm{nat}}$K~\cite{DDEPWG}, $A_{\mathrm{K}} = 39.0983(1) $ is the atomic
weight of potassium~\cite{Audi:2002rp};
$c_{\mathrm{K}} = 20$ ppb corresponds to the upper limit as reported
by DAMA~\cite{Bernabei:2008yh}.
The decay will be perceived as a mono-energetic event at the $\Arforty$
K-shell binding energy of $E_K=3.2\,\keV$.  The signal shape is hence
dominated by the energy resolution of the detector $\sigma(\keV) =
0.448 \sqrt{E} + 0.0091 E$~\cite{Bernabei:2008yh} where $E$ is in units of keV. We find good
agreement with the observed shape of the background and the total rate
in the energy bin $[E_{\mathrm{min}}, E_{\mathrm{max}}]$ then reads
\begin{align}
\label{eqn:gsECbackgr}
B_{\mathrm{EC}} =  \frac{\Gamma_{\mathrm{EC}} }{ 2 }
\left[ \erf \left( \frac{E_{\mathrm{max}}-E_{K}}{\sqrt{2} \sigma }
  \right) - \erf \left( \frac{E_{\mathrm{min}}-E_{K}}{\sqrt{2} \sigma
    } \right) \right] .
\end{align}

The $\beta^{-}$ decay to \Caforty\ , contribution (ii), is also easy
to account for since the emitted electron is entirely contained in the
crystal where the decay happened. Thus, the spectrum is found from the
3U shape factor for \Caforty.%
\footnote{
  The $\beta^{-}$ spectrum shown in Ref.~\cite{Kudryavtsev:2010zza} is
  based on a GEANT4 simulation which erroneously computed the shape
  factor of an allowed decay. In
  Fig.~\ref{fig:dama_low_energy_spectrum} we have employed the correct
  shape factor. It mainly affects the kinematical shoulder at
  1311~keV.}
At low energy this is a fairly flat background as shown in
Fig~\ref{fig:dama_low_energy_spectrum}. The shoulder present near the
kinematical end-point of 1311 keV might offer a straightforward way to
estimate the level of \Kforty\ contamination level if it can be
clearly observed above the other backgrounds. Such a determination would obviate the need to rely on MC for the purpose of determining the concentration of $^{\mathrm{nat}}$K. Our results
indicate that given the background levels from other sources (mainly
$^{238}$U and $^{232}$Th), this shoulder could be observed or a useful
upper bound can be obtained. Unfortunately, the DAMA collaboration has
not released the spectrum at this energy range. Assuming that the bump at 3 keV is entirely due to background we predict that a shoulder with a height of about $0.04$~cpd/kg/keV should be seen at around 700 keV, associated with the $\beta^{-}$ decay of \Kforty\ .

Finally, a MC simulation is necessary to estimate the rate
associated with (iii) where the 1460 keV gamma-ray escapes entirely
undetected. This was done independently in
Ref.~\cite{Kudryavtsev:2010zza} where the total contribution from (i),
(ii) and (iii) was taken into account with
${\rm BR}_{\mathrm{EC}}=0.2~\%$. In order to investigate the effect of
different values of ${\rm BR}_{\mathrm{EC}}$ we subtracted the
contribution of (i) and (ii) from the total
\Kforty\ spectrum quoted in Ref.~\cite{Kudryavtsev:2010zza} to obtain
the spectrum associated with the EC decay to the excited state,
contribution~(iii). From Fig.~\ref{fig:dama_low_energy_spectrum} we
find that (iii) is indeed the dominant source of $\Kforty $ background
at $E_R\sim 3\,\keV$ by a factor of about $5\times
(0.2/{\rm BR}_{\mathrm{EC}})$. Given the importance of this background it would be interesting to compare this
number with DAMA-internal MC simulations.

We close this section by noting that the flat spectrum at low energies
associated with the 3U transition to \Caforty\ is a universal feature
of $\beta^{-}$ decay. In the non-relativistic limit, the product of
Fermi function and phase space volume in $\beta^{-}$ decay reads,
\begin{align*}
  \frac{2\pi \eta}{1-e^{-2\pi \eta}} \times p^2 (E_0 - E)^2 dp \to
  const \times dE \quad (\eta \gg 1),
\end{align*}
where $\eta = Z\alpha /v$ is the Sommerfeld parameter and $v$ is the
asymptotic electron velocity; for $\beta^{-}$ decay of $\Kforty$, $\eta
>1$ for $E \lesssim 200\,\keV$ and the last equality is approximately
constant in this energy regime.
We suggest that the essentially flat low-energy event rate seen in
DAMA (more than 20 data-points reported up to $E_R \leq 10\,\keV$ and
between $20-30\,\keV$ with negligible error-bar) is strongly
suggestive of the presence of such a constant background component
(\textit{e.g.}  through an unaccounted $\beta$-emitter or from
low-energy Compton background) at a level of%
\footnote{An extended discussion regarding this hypothesis can be
  found in an addendum to this paper~\cite{Pradler:2012bf}.}
\begin{align}
\label{eq:Bflat}
  B_{\mathrm{flat}} \simeq 0.85~\cpd/\kg/\keV .
\end{align}
In what follows, we will assume that this is indeed the case and work
out the implications for a successful DM interpretation given that
potassium is present as well. Hence, $B_0 = B_{\mathrm{flat}} + B_{40}$
where $B_{40}$ is the low-energy background from $\Kforty$.

\section{DM signal interpretation in the presence of backgrounds}
\label{sec:sign-interpr-pres}

We now proceed to investigate the influence of backgrounds on the
DAMA DM signal claim.  
To better quantify the issue let us write $R_0 = B_0 + S_{0} $ where
$B_{0}$ is background and $S_0$ is the unmodulated contribution of
any tentative DM signal. These quantities depend on \Er\ but are
assumed to be stationary in time. The total event rate in the presence of
a DM signal is then given by\footnote{See Ref.~\cite{Freese:2012xd}
  for a review and Ref.~\cite{Chang:2011eb} for a discussion of the
  general temporal variations expected.}
\be
\label{eq:rate}
R(t) = B_0 + S_0 + S_m \cos{\omega (t-t_0)} .
\ee
The observed modulating fraction of the event rate has its maximum at
a recoil energy of approximately $3\,\keV$, and in the
$(2-4)$\,\keV\ energy bin,
\be s^{\mathrm{obs}}_m = \left. \frac{S_m}{B_0+S_0}\right|_{
  2-4\,\keV} \simeq 2\%.  \ee
Written in this form, it becomes clear that an understanding of the
background $B_0$ is essential for a successful interpretation of the
DAMA results.

A viable DM model must reproduce the observed modulation amplitude,
and in particular have sufficiently large modulation amplitude near
$3~\keVee$ recoil energy. Given a certain background level $B_0$ we have
\be
\label{eqn:smax_theory}
s^{\mathrm{max}}_m \ge \left(1 + \frac{B_0}{S_0} \right)
s^{\mathrm{obs}}_m \approx 2\% \times \left(1 + \frac{B_0}{S_0},
\right) \ee
where $s^{\mathrm{max}}_m \equiv S_m/S_0|_{ 2-4\,\keV}$ is the maximum modulation fraction.
For very low background levels ($B_0/S_0 \ll 1$) the requirement on
the signal's modulation amplitude $s^{\mathrm{max}}_m \ge 2\%$ can be
easily satisfied by most models of DM that aspire to explain the DAMA
result. However, considering the bump around 3~keV in the unmodulated
rate, it is almost certainly the case that the background levels are
not that low.

\begin{figure}[tb]
\begin{center}
\includegraphics[width=1\columnwidth]{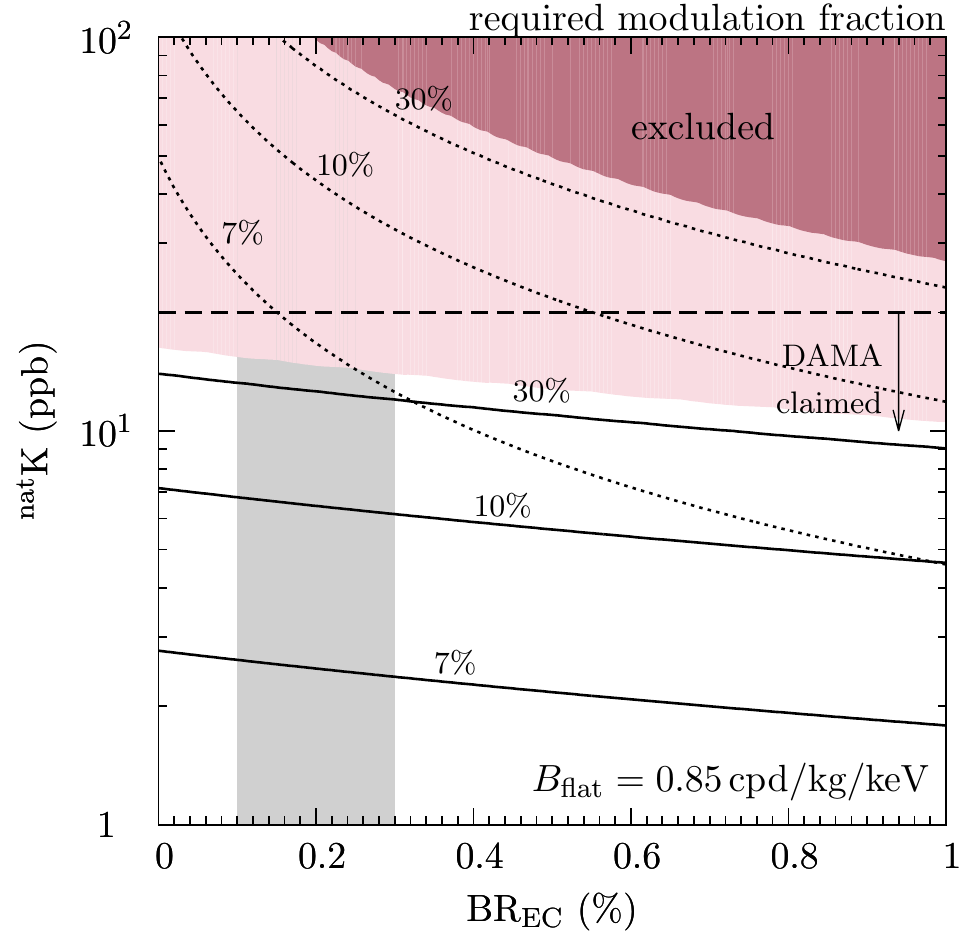}
\end{center}
\caption{Required modulating fraction $S_m/S_0$ of a DM signal at
  $3\,\keVee$ nuclear recoil energy in the presence of a flat
  background and potassium contamination. Solid (dotted) lines are the
  associated contours in the parameters ${\rm BR}_{\mathrm{EC}}$ and
  $c_K$ with (without) background contribution (iii) as detailed
  in Sec.~\ref{sec:bkg}. In the (red) shaded regions, the required
  modulating fraction exceeds 100\%. The vertical gray band indicates
  the nominal value for $\BREC$ and its uncertainty as quoted in Eq.~(\ref{eq:ECtoBetaPlus}). The horizontal dashed line shows the DAMA claimed upper
  limit on $\Kforty$ contamination. }
\label{fig:POT_contours}
\end{figure}

Using the background contributions just discussed,
including the flat contribution in Eq.~(\ref{eq:Bflat}), we compute the required modulation
fraction in $(2-4)\,\keV$ from Eq.~(\ref{eqn:smax_theory}) as a function
of both the branching ratio ${\rm BR}_{\mathrm{EC}}$ and the contamination level of $^{\rm
  nat}$K; The unmodulated signal rate, $S_0$, is required to explain the
observed single hit rate in that bin, $R_0 \simeq 2.2\,\cpd/\kg$.
Contours of the required fraction are shown in
Fig.~\ref{fig:POT_contours}.
Since the contribution from EC into the excited
state of \Arforty\ (case (iii) in the previous section) is obtained from MC and its uncertainty is unknown we consider two scenarios: fractions including this contribution are shown as solid contours whereas fractions neglecting this contribution are shown as dotted contours. In the
respective shaded regions $s^{\mathrm{max}}_m$ would be larger than
100\% and these regions are therefore excluded.

As is clear from the figure, any model with small modulation fraction
of a few \% is already strongly disfavored by the data. Light DM models, which usually predict $\sim 10\%$ modulation
fraction (see e.g.~\cite{Kopp:2011yr}), are also in tension with the data. The viability of these models can be better assessed with a measurement of
${\rm BR}_{\mathrm{EC}}$ and a more thorough investigation of the
contamination levels from $^{\rm nat}$K.
Finally, inelastic DM models~\cite{TuckerSmith:2001hy} such as MiDM~\cite{Chang:2010en} can yield a large modulation
fraction ($\gtrsim 30~\%$) due to the heightened sensitivity to the
galactic escape velocity. Such models are unlikely to be ruled out
through our simple considerations.
It is however remarkable that if the DAMA quoted bound on
$^{\mathrm{nat}}$K is indeed saturated ($c_K = 20~$ppb) and the
strength of background (iii) is adequately caught by the MC simulation,
then the required modulation fraction cannot be attained by any DM
model. The necessity of a more comprehensive discussions of
backgrounds in the signal region by the DAMA collaboration is evident.

\section{Comments on realizing a measurement of the rare decay of
  $\Kforty$}

Given the above, a direct measurement of the branching ratio of the rare electron capture decay of \Kforty\ directly to the ground state of \Arforty\ is clearly desirable. 
The very low energy ($\sim 3\keV$) released make such a measurement challenging and higher concentrations of \Kforty\ compared with the natural abundance is likely needed. 
It is possible to obtain enriched potassium samples with a $\Kforty$
concentration of $14\%$ and higher~\cite{TSC:2012}. Such
a high concentration would lead to an activity at the level of
\begin{align}
  \sigma_{\rm K} \approx 380 {\rm~ day^{-1}\,mg^{-1} } \times
  ({\rm BR}_{\mathrm{EC}}/0.2\%), 
\end{align}
so that statistics is unlikely to be the limiting factor of any effort
to measure ${\rm BR}_{\mathrm{EC}}$.
The main difficulty in measuring the rare decay directly to the ground
state is the small branching ratio compared with the more common EC
decay $\Kforty \rightarrow \Arforty^{*}(1460)$. As discussed before,
if the 1460~keV $\gamma$-ray escapes the detector, only the
deposition of $3~\keV$ of energy is registered, which mimics the direct decay. However, with an additional surrounding
\textit{anti-coincidence} veto, this background becomes reducible.
Given that the ratio of branching ratios of the two decays is $\sim
0.2\%/10\% = 2\%$ the veto efficiency only needs to be somewhat better
than $1\%$. Whether this is a realistically attainable efficiency is
beyond our expertise to determine.

The low energy release of $\sim 3~\keV$ is not an easy energy range
for detection, but NaI(Tl) crystals such as the ones used by the DAMA
collaboration have been demonstrably sensitive in this
range. Germanium based detectors such as the ones employed by the
CoGeNT collaboration~\cite{Aalseth:2008rx} have an even lower
threshold and better energy resolution. A precise measurement of the
electron capture decay of $\Kforty$ seems possible. Finally, we note
that scintillating crystals can also be grown from KI(Tl) powders and
have in fact been used in measurements of the $\Kforty$ decay
scheme~\cite{PhysRev.79.940}. Because of the relatively smaller
scintillation light output in comparison with NaI(Tl), it remains to
be proven if sensitivity at 3~keV can be attained.

Finally, there has recently been some renewed interest in the exotic
possibility of temporal variations in nuclear decay
rates~\cite{Jenkins:2008vn}. The data used is rather old and suspect,
and there has since been several refutations of these
claims~\cite{Cooper:2008kj,Norman:2008gz, Hardy:2011ku,
  Bellotti:2012if}. However, we point out that no conclusive exclusion
of such an effect has been reported in the case of EC decay rates.%
\footnote{Ref.~\cite{Norman:2008gz} searched for correlation between
  the Sun-Earth distance to the ratio of $^{22}{\rm Na}/^{44}{\rm
    Ti}$. $^{44}{\rm Ti}$ decays via EC, whereas $^{22}{\rm Na}$ does
  so only $\approx 10\%$ so in principle this ratio can be sensitive
  to variations associated with EC.  A power-spectrum analysis of the
  data (which~\cite{Norman:2008gz} did not perform) seems to reveal
  significant power at a period of one year and therefore such
  variations cannot be robustly excluded.}
Thus, it might be of some interest to search for such modulations in
the electron capture decay of $\Kforty$ directly to the ground state
of $\Arforty$. This is especially interesting given the unique nature
of this rare decay and its direct relevance to the annual modulations
claimed by the DAMA collaboration.

\section{Conclusions}
\label{sec:disc-concl}

In this letter we have identified a branch of \Kforty\ decay which is yet experimentally unverified and whose strength is only
estimated from theory.
Aside from its intrinsic importance as the only such known decay of its kind, this branch has important ramifications for DM direct detection. The
in-situ presence of \Kforty\ is well established in the DAMA detector although the precise contamination level is not clear. Interestingly, it presents a background in the very signal region from where
the collaboration derives its claim for DM detection.
Before closing we would like to highlight some of our findings:
\begin{itemize}
\item The $\Kforty$ EC decay to the ground state of $\Arforty$---a
  third forbidden unique transition and the only one known of its
  kind---lacks a dedicated measurement to-date.
\item Nuclear data evaluations use extrapolations to infer the
  associated branching. Although we find good agreement when using
  leading-order theory, a direct experimental verification is
  nevertheless much desired and called for.
\item Depending on the actual concentration of $\Kforty$, a DM
  explanation of the DAMA signal based on elastic scattering and a
  Mawellian halo velocity profile may already be excluded.
\item With a potassium concentration of 10~ppb (which is below the
  DAMA inferred upper limit), its $\beta^{-}$ decay may very well
  dominate the spectrum at 1~MeV. The 3U shape with its kinematical
  shoulder could then be used as an independent measurement of the
  $\Kforty$ concentration.
\end{itemize}
Based on the above findings we propose the following steps to help improve the situation:
\begin{itemize}
\item A dedicated measurement of the $\Kforty $ EC decay into the
  ground state---potentially over an extended period of time to exclude the possibility of temporal variations---is
  itself a missing piece in the experimental verification of leading
  order weak decay calculations and will help to settle the role of
  this decay in the DAMA experiment.
\item We propose that the concentration levels of $^{\mathrm{nat}}$K
  can be inferred from the activity level associated with the
  $\beta^-$ shoulder of \Kforty\ at around 1 MeV. This method has been
  verified impressively in~\cite{Cebrian201260} for a NaI prototype
  crystal by the ANAIS collaboration.
  A count rate in DAMA much greater than $0.04$ cpd/kg/keV will severely
  undermine a DM interpretation of the signal. This assumes that the
  other reported concentrations are reliable and that a flat
  background at a level of $0.85\,\cpd/\kg/\keV$ is present.
\item The DAMA spectrum between 10-20~keV should be released as it 1)
  helps to clarify the hypothesis of a flat $\beta^{-}$ backgrounds and 2) potentially allows to identify
  further EC elements if other K-shell capture ``bumps'' were present. 
\item We urge the collaboration to provide more details concerning the probability of coincidence
  events when the 1460 keV gamma-ray from $\Kforty$ decay
  escapes one crystal but is detected elsewhere. A comparison with the
  independent study by~\cite{Kudryavtsev:2010zza} could yield precious
  insights into the reliability of the MC models and corroborate the
  determination of $^{\mathrm{nat}}$K through the coincidence method.
\item Likewise, showing the $\alpha$-peaks in the spectrum above $\sim
  2~\MeV$ could reaffirm the average levels of $^{238}$U and
  $^{232}$Th determined by DAMA and clarify their contributions to the
  count rates in the DM signal region.
\end{itemize}

\begin{acknowledgments}
  We acknowledge helpful discussions with K.~Blum, A.~Chen, R.~Lang,
  V.~Kudryavtsev, F.~Pr\"obst and N.~Weiner. BS and IY are supported
  in part by funds from the Natural Sciences and Engineering Research
  Council (NSERC) of Canada. Research at the Perimeter Institute is
  supported in part by the Government of Canada through Industry
  Canada, and by the Province of Ontario through the Ministry of
  Research and Information (MRI).
\end{acknowledgments}

\bibliography{K40bib}

\end{document}